\documentclass[prb,twocolumn,showpacs,floats,eqsecnum,amsmath,amssymb]{revtex4}
\usepackage[dvips]{graphicx}

\begin{document}


\title {Effect of synaptic plasticity in the structure and dynamics of disordered networks of coupled neurons}
\author {M. Bayati and A. Valizadeh}

\affiliation{Institute for Advanced Studies in Basic Sciences, P.O. Box 45195--1159, Zanjan, Iran}


\begin{abstract}
In an all-to-all network of integrate-fire oscillators in which there is a disorder in the intrinsic firing rates of the neurons, we show that through spike timing-dependent plasticity the links which have the faster oscillators as presynaptic, tend to be strengthened while the links originated from the slow spiking neurons are weakened. The emergent effective flow of directed connections, introduces the faster neurons as the more influent elements in the network and facilitates synchronization by decreasing the synaptic cost for onset of synchronization. \end{abstract}

\vspace{2mm} 

\maketitle

Experimental studies indicate that excitatory synapses are very sensitive to the temporal order of firing of pre- and postsynaptic neuron\cite{stdp1}. A synaptic efficacy with spike timing-dependent plasticity (STDP), is found to increase if firing of a presynaptic neuron occurs in advance of firing of a postsynaptic neuron, and to decrease if the temporal order of firing is reversed\cite{stdp1,stdp2}. Indeed, STDP is widely thought to underlie learning processes, and in itself constitutes a broadly interesting phenomena\cite{stdp3,stdp4}.

The vast studies on synchronization both in the scale of few neurons and in large networks, reveals that with STDP neural synchronization is more rapid and robust\cite{zigul}. Comparing to networks of fixed coupling strength, in the networks in which the couplings change according to STDP, the regions of synchronization in the parameter space are wider, e.g. they can suffer larger mismatch in intrinsic frequencies yet showing synchronized behavior.


 While most of the early studies on synchronization properties of the complex networks ignore the evolution of network structure and the directionality of the links, recent studies address both the effect of the links directionality\cite{directed1,directed2,directed3,directed4} and time-dependent coupling strengths\cite{evolutionary}. When networks are directed, the Jacobian or Laplacian matrices will have complex eigenvalues which influences both the stability\cite{nature} and the dynamical organization of complex networks\cite{directed1,directed4}. Here we study how STDP changes the structure of the directional links of a neuronal network, in an initially (topologically) homogeneous network consisting of nonidentical oscillators. Starting with an all-to-all network with symmetric couplings, we will show that disorder in the intrinsic firing rates leads to asymmetric couplings in a predictable manner, i.e. the evolution of the network is such that the influence of the neurons (strength of outgoing couplings) with higher rate of activity is enhanced and in turn, the strength of incoming coupling to the slow neurons is increased. Then we show the coupling cost for the onset of synchronization for such a network, which has an effective network flow of the directed connections from fast to slow components, is smaller than that of a symmetric network\cite{directed4}. So in a network of constant sum of the node strengths, such effective flow of connections leads to more organized dynamics. In turn we show the evolution of the synaptic strengths in the network depends on whether or not synchrony is achieved through STDP.

The model network consists of $N$ pulse-coupled non-identical oscillators, each of them defined by a linear first order equation:
\begin{equation}\label{eq1}
    \tau \frac{dv_{j}}{dt}=-v+I_{j}+I_{ij},
\end{equation}
in which $v_{j}$ is a voltage like variable for each neuron labeled by $j=1,2,...,N$, $\tau=1$ is the time constant. Every time a threshold value $v_{th}=1$ is touched, neuron {\it fires} and the voltage resets to $v_{res}=0$. $I_{j}$ is the external excitation (current) and $I_{ij}$ is {\it synaptic current} with the neurons $i$ and $j$ as the pre- and post-synaptic neurons, respectively. The spike are recorded by the {\it neuron response function}\cite{dayan} defined as $x_{j}(t)=\sum_{m} \delta(t-t_{j}^{m})$ where $t_{j}^{m}$ is $m^{th}$ time when the neuron $j$ fires and $\delta(x)$ is the Kronecker delta function. The synaptic current $I_{ij}$ is defined as
\begin{equation}\label{eq2}
    I_{ij}=a_{ij} g_{ij} x_{i}(t),
\end{equation}
where $a_{ij}$ is the element of the adjacency matrix\cite{network} which is one when there is a direct connection between the neurons $i$ and $j$ as the pre- and post-synaptic neurons and zero otherwise. With $a_{ij}=1$ neuron $j$ receives a kick by the strength $g_{ij}$ every time neuron $i$ fires. Synaptic strength $g_{ij}$ is positive throughout this study to model excitatory synapses. For later convenience we call the matrix formed by the elements $a_{ij} g_{ij}$ \emph{weighted adjacency matrix}.

With the minimal model we used, with the equal time constants of the neurons, inhomogeneity in the intrinsic activity rates is imposed by choosing the external currents $I_{j}$ from a distribution, note that this could also be imposed by choosing neurons with different time constants and equal feeds.  Distribution of the firing rates can then be calculated using the relation of the firing rate of a LIF neuron to input current as $r=[\tau \ln(I/(I-v_{th}))]^{-1}$.

The time-dependent synaptic coupling strength $g_{ij}$ changes depending on the dynamics of the presynaptic and postsynaptic neurons.
Through STDP $g_{ij}$ changes by $\Delta g_{ij}$, which is a function of the time difference $\Delta t=t_{j}-t_{i}$ between the times of postsynaptic and presynaptic spikes. Synaptic modification $\Delta g_{ij}$ is provided by
\begin{equation}\label{eq3}
   \Delta g_{ij}=A_{\pm} sgn(\Delta t) exp(-|\Delta t|/\tau_{\pm}),
\end{equation}
where the parameters $\tau_{+}$ and $\tau_{-}$ determine the ranges of pre-to-postsynaptic interspike intervals over which synaptic strengthening and
weakening occur. $A_{+}$ and $A_{-}$, which are both positive, determine the maximum amounts of synaptic modification which occur when $\Delta t$ is close to zero\cite{stdp2}. $A_{+}$ $(A_{-})$ and $\tau_{+}$ $(\tau_{-})$ are used when $\Delta t$ is positive (negative). Since the additive STDP is used, divergence of the synaptic strengths is prevented by assuming limiting values for the synaptic strength.

It is noted before by Gilson {\it et al} that symmetry of the coupling matrix is broken by STDP\cite{gilson1,gilson3}. We study how is the possible effect of such asymmetry on the dynamics of the network and how the dynamics in turn affects the structure of the connections in the network. We first define the \emph{synaptic cost} as the sum of the all synaptic strengths in the array. To quantify the asymmetry, we define {\it link imbalance} as the difference of the synaptic strengths between two nodes $C_{ij}=-C_{ji}=a_{ij}g_{ij}-a_{ji}g_{ji}$. Furthermore, we introduce the \emph{strength} of the node as the sum of the all the incoming synaptic strengths, the synapses which have the neuron $j$ as the post-synaptic $C^{-}_{j}=\sum_{i}a_{ij}g_{ij}$; and \emph{sensitivity} of node as the outgoing synaptic strengths, those which have the neuron $j$ as pre-synaptic $C^{+}_{j}=\sum_{i}a_{ji}g_{ji}$. This sums can also be interpreted as the sum of the elements of $j$th column and $j$th row of the weighted adjacency matrix, respectively. We call the difference between the outgoing and incoming synaptic strengths for each neuron $C_{j}=C^{+}_{j}-C^{-}_{j}$ as the {\it node imbalance}. A positive node imbalance means the neuron's outgoing synapses are stronger than its incoming synapses and vice versa. Also introducing {\it network imbalance} as $C_{net}=1/N^2 \sum_{i,j} sign(i-j) C_{ij}$, we can deduce that the mean strength of the fast neurons is larger than their mean sensitivity if $C_{net}$ is positive and vice versa. We will show later that STDP can increase network imbalance but before, we inspect effect of predetermined network imbalance on the dynamic of a network with static synapses.

\begin{figure}[ht!]
\vspace{0cm}\centerline{\includegraphics[width=11cm]{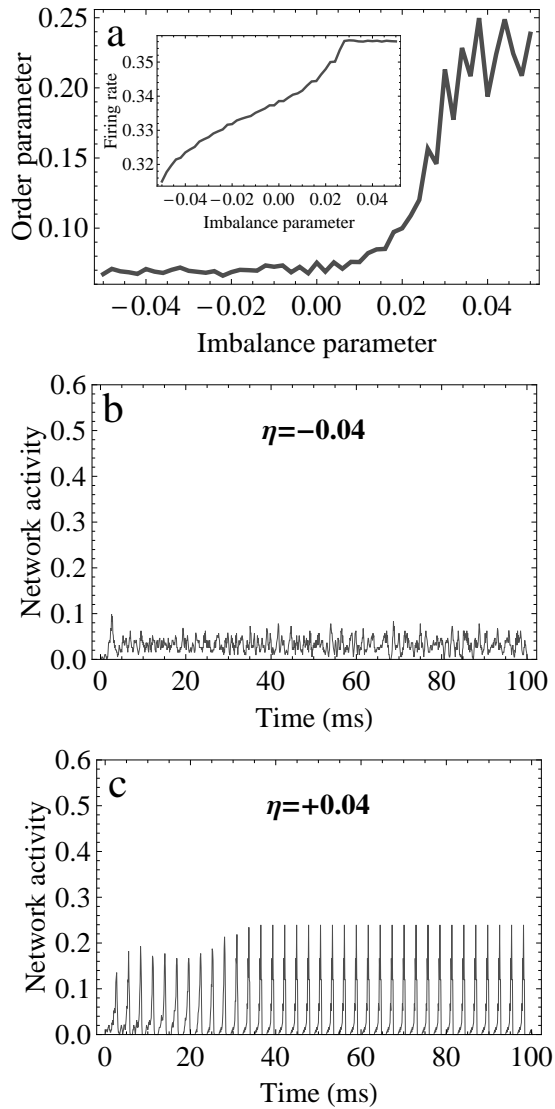}}

\vspace{-0cm} \caption{(a) The order parameter, the amplitude of the network activity as a function of parameter of imbalance. In a network of $N=64$ neurons with all-to-all connections, input currents are chosen as $I_{j}=1+0.001j$ and synaptic strengths as $g_{ij}=1/N[.08+\eta\hspace{.1cm}sign(i-j) f(|i-j|)]$ with $ f(x)=x^{(1/\sqrt{N})}$. The range of $\eta$ we have studied is such that all the synaptic strengths are positive. Increasing $\eta$ beyond this range results in negative synaptic strengths which we do not consider here. In (b) and (c) the network activity is shown for two different values of imbalance parameter: periodic behavior of the network activity with relatively large amplitude in (c) indicates synchrony of the neurons. Inset of (a) shows mean firing rate of the neurons in the network vs. parameter of imbalance. }\vspace{0cm} \label{fig1}
\end{figure}

We construct a fully connected network ($a_{ij}=1$ for every $i$ and $j$)  with the link imbalance as a variable parameter, assuming the synaptic strengths as $g_{ij}=1/N[g_{0}+\eta\hspace{.1cm}sign(i-j) f(|i-j|)]$, with constant $g_{0}$ and $f(\xi)$ a monotonically increasing function of $\xi$. Then the link imbalance $C_{ij}=2\eta\hspace{.1cm} sign(i-j) f(|i-j|)$ and the network imbalance can be controlled by the parameter $\eta$. The external currents are chosen equally spaced in the interval $[I_{0}-\delta,I_{0}+\delta]$, and the neurons are labeled in order of increasing input current, i.e. the $j=1$ neuron has the smallest input and so on. There are two points worth noting: first since the nodes are labeled in order of increasing intrinsic firing rates, the two neurons with larger difference in intrinsic firing rates have a link with larger imbalance. Second, sum of the all synaptic strengths in the network remains constant (equal to $g_{0}$) when changing imbalance parameter $\eta$.

Now we inspect how the dynamics of the network is affected by changing imbalance parameter. The \emph{network activity} is defined as the average response functions of all the neurons in the array $X_{net}(t)=1/N\sum_{j}x_{j}(t)$. Inphase firing of the large fraction of neurons in the array leads to oscillatory behavior of the network activity function with large amplitude, so the amplitude of the network activity function can be used as an order parameter showing how synchronized are the firing of the neurons in the network. In Fig. 1a we have shown how the order parameter changes when we increase the imbalance parameter in a network with constant sum of the couplings. The plots show that the neurons can be synchronized when we increase the strength of fast neurons and decrease their sensitivity. It is also shown negative imbalance has no effect on the coherence of the behavior of the neurons, i.e. they are outgoing synapses from the fast neurons which should be strengthened to achieve synchrony. It is also shown in the inset of Fig. 1a that mean firing rate of the array increases with imbalance parameter, which is a reasonable consequence of the increase of strength of faster neurons. In such a system synchrony can be interpreted as the triumph of the fast components to dictate their dynamics on the slower neurons; increase in the strength of the fast component (meanwhile the sensitivity of the slow components increases) leads to increase of both the mean firing rate of the network and the degree of synchrony.

\begin{figure}[ht!]
\vspace{0cm}\centerline{\includegraphics[width=13cm]{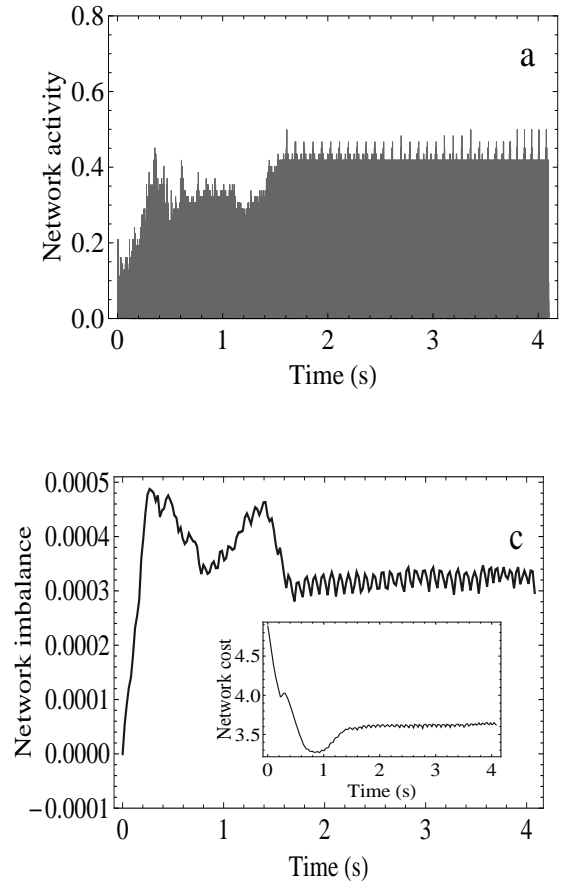}}

\vspace{-5cm} \caption{ In (a) the network activity is shown when the synaptic strengths evolve through STDP. The boosted amplitude of activity is due to transition of the network to oscillatory state which indicates synchrony in the array. In (b) evolution of the network imbalance is shown. Inset shows the evolution of the network cost, sum of the all the synaptic strength in the network. The currents are chosen as $I_{j}=1+0.001j$, lower and upper cutoffs are zero and $0.16/N$, respectively with the network size $N=64$. The initial couplings are all equal to $0.08/N$. Parameters of STDP are $A_{+}=0.000055$, $A_{-}=0.000050$, $\tau_{+}=10$ and $\tau_{-}=15$.  }\vspace{0cm} \label{fig1}
\end{figure}

 We now let the synaptic strengths to evolve through STDP, and investigate how the initial synaptic strengths and consequent possible organized dynamics of the network can affect the evolution of the structure of the network. Again we consider a fully connected network with initially equal symmetric synaptic strengths. We assume antisymmetric STDP profile with usual criterion ($A_{+}>A_{-}$ and $A_{+} \tau_{+} < A_{-}\tau_{-}$) with zero lower cutoff. We examine two situations: in both of them the initial synaptic strengths are not enough to overcome disorder in the array and the neurons are unsynchronized when STDP is absent. Asymmetry induced by STDP in one of the experiments leads to synchrony where in second experiment the neurons remain unsynchronized in the steady state as is shown in Figs. 2 and 3.

When STDP leads to synchronized firing of the neurons (Fig. 3), a net synaptic flow is constructed from the fast to the slow neurons, which is reflected in the value of network imbalance as it takes positive value in the steady state. The positive network imbalance indicates most of the weakened synapses are those from slow to fast neurons and most of those which are strengthened, are from faster neurons to slower ones. With the parameters we have chosen the synaptic cost of the network decreases; this is of great importance since synchrony is achieved despite of such a decrement in the synaptic cost. This is consistent with the above result which imbalance lowers the synaptic cost for onset of synchronization. We mention here that the evolution of synaptic cost is dependent to the choice of parameters of STDP and with a minor change of parameters, synaptic cost may increase. But nevertheless, when the final state of the network is synchronized network imbalance increases. We note here that although asymmetry induced by STDP has been reported before\cite{gilson3}, with the differences in intrinsic rate of firing of the neurons, asymmetry is established such as a net structural flow of the weighted links (from fast to slow neurons) is created in the network. In the other experiment in which smaller values are chosen for the initial synaptic strengths (while keeping the upper cutoff unchanged), the plasticity can not lead to synchrony (Fig. 3). In this case the time course of the network imbalance is dependent to the initial condition and it can choose both positive and negative values.

\begin{figure}[ht!]
\vspace{0cm}\centerline{\includegraphics[width=11cm]{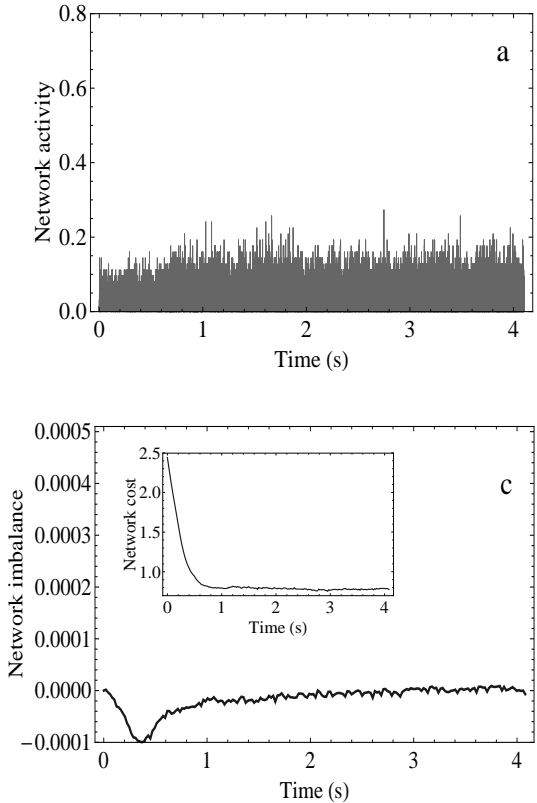}}

\vspace{-4.2cm} \caption{ The network activity (a), the evolution of some of the synaptic strengths (b) and the evolution of the network imbalance are shown for the same network as Fig. 2 with smaller initial synaptic strengths. All the parameters are the same as Fig. 2 except for the initial couplings which here are all equal to $0.04/N$. Inset shows the evolution of the network cost, sum of the all the synaptic strength in the network. }\vspace{0cm} \label{fig1}
\end{figure}

We have repeated a similar experiment with the Hodgkin-Huxley neurons with chemical synapses (see appendix), to inspect whether the results are applicable in the more biologically plausible models. As is shown in Fig. 4, role of STDP is to decrease the effect of discrepancy in the intrinsic firing rates and organize the dynamics of the neurons. In turn, emergent structure of the network is shown in Fig. 5 where a nearly triangular weighted adjacency matrix is formed and network imbalance is reasonably increased.

\begin{figure}[ht!]
\vspace{0cm}\centerline{\includegraphics[width=8cm]{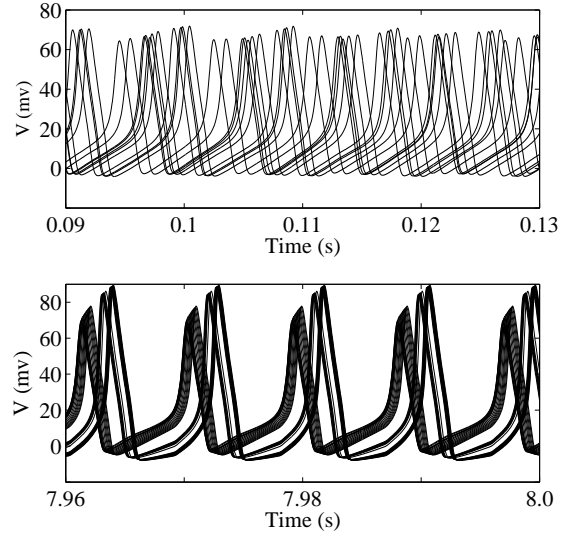}}

\vspace{0cm} \caption{Membrane voltages of the sample neurons is given in two different times in initial (upper plot) and steady state (lower plot). the parameters of the HH neurons and synapses are given in the appendix. Other parameters are $A_{+}=9ns$, $A_{-}=8.6ns$, $\tau_{+}=20ms$ and $\tau_{-}=30ms$.}\vspace{0cm} \label{fig2}
\end{figure}

\begin{figure}[ht!]
\vspace{0cm}\centerline{\includegraphics[width=9cm]{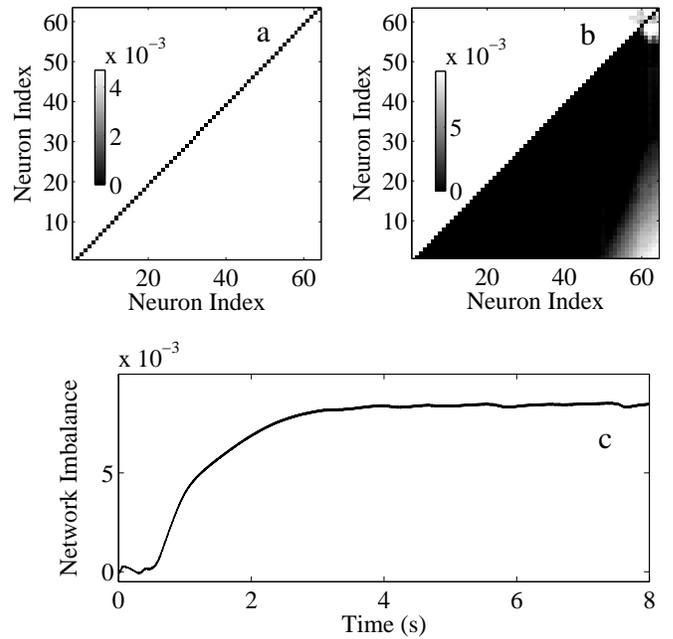}}

\vspace{-0cm} \caption{ In a network composed of Hodgkin-Huxely neurons and chemical synapses, the initially symmetric the weighted adjacency matrix (a), evolves to a nearly triangular matrix in the steady state (b). In (c) the evolution of network imbalance is shown. All the parameters are those of Fig. 4.}\vspace{0cm} \label{fig1}
\end{figure}


In passing we show that for two weakly connected, non-identical neurons with excitatory couplings, both increasing the strength of the fast neuron and decreasing its sensitivity enhances domain of synchronization. In other words such asymmetric connection can lower the threshold for onset of synchronization. We consider two neurons with the inputs $I_{2}=I_{1}+\delta$ which are connected by two directed couplings with the strengths $g_{12}$ and $g_{21}$. With positive mismatch parameter $\delta >0 $ the second neuron is the faster. Looking for existence criterion for the inphase $1:1$ synchronization, we consider the two cases in which one of the neurons (master) fires and makes also the other neuron (slave) to fire. In first case we assume the faster neuron first fires at time $t_{i}$ and the slower fires just after it i.e. the fast neuron is master and the slower is slave. With no refractory period firing of the slave neuron changes the voltage of the master by $g_{12}$ and the fast neuron would fire again at time $t_{i+1}=t_{i}+\ln(\frac{I_{2}-g_{12}}{I_{2}-1})$. If at the time of firing of the master neuron the voltage of the slave neuron is larger than $1-g_{21}$, it fires following the fast neuron, that is
\begin{equation}\label{eq1}
    I_{1}>(1-g_{21})\frac{I_{2}-g_{12}}{1-g_{12}}.
\end{equation}
Since we have assumed here the slave neuron is the slower one, it can not exceed the fast neuron and the above equation solely determines the existence condition for the inphase solution. Assuming $I_{1}=I_{2}-\delta$ this equation determines the maximum mismatch which allows inphase $1:1$ synchronization. The second case which as we will see is possible for large values of coupling constants, assumes that the slower neuron is the master. In this case a criterion similar to Eq. 4 exists and also we should prevent the faster neuron to exceed the slower neuron i.e. at the time of next firing of the master neuron $t_{i+1}$ the voltage of the fast neuron should be less than threshold. Putting together we get
\begin{equation}\label{eq1}
    \frac{I_{1}-g_{21}}{1-g_{21}}>I_{2}>(1-g_{12})\frac{I_{1}-g_{21}}{1-g_{21}}.
\end{equation}


In the equations above two points are worth noting: for small values values of coupling constants which is matter of our study, just the synchronized state with fast neuron as the master can exist and this state can not be achieved with large values of strength of the slow neuron $g_{12}$. In other words $g_{21}$ appreciates synchrony and $g_{12}$ opposes it when the state in which the fast neuron is dynamically master, is the only possible inphase state, that is, for small values of synaptic strengths. Although it can be shown for near threshold currents $I_{i} \sim 1+O(\epsilon)$ and small synaptic strengths $g_{ij}\sim O(\epsilon)$, effect of the $g_{12}$ is of order $\epsilon^2$, but for larger input currents the effect of the strength of slow neurons can be comparable with that of fast neuron. With STDP for two weakly connected neurons, our results show that the strength of the fast neuron always increases and that of the slow neuron decreases and as noted above both of them appreciate synchrony. When synchrony is achieved (with the fast neuron as the master), the rate of change of the synaptic strengths increases and they are then just limited by the cutoffs considered in the model.

To conclude, we have shown that in the systems of weakly connected neurons with excitatory synapses, when there is a mismatch in the intrinsic firing rates of neurons, a special asymmetric arrangement of synaptic constants can enhance synchrony. In this arrangement directed links from the faster elements to the slower ones should be stronger. In a two neuron system, this result is verified by a simple analytic reasoning. We have also showed that spike timing-dependent plasticity in such disordered networks, can organize the firing of the neurons by imposing such asymmetry on the matrix of synaptic strengths. In turn, the emergent structure of the synapses in the presence of STDP depends on weather or not synchrony is achieved in the network.
\appendix
\section{The Hodgkin-Huxley model and chemical synapses}

The membrane voltage of the neuron in the Hodgkin-Huxley (HH) model is described by\cite{HH}:
\begin{equation}
c\frac{dv_{j}}{dt}+I_{na}+I_{k}+I_{l}+I_{ij}=I_{j}.
\end{equation}
$c$ is the capacitance per unit area of the membrane which is taken as $1\mu F/cm^{2}$ and $I_{j}$ stands for the external current. $I_{l}=g_{l}(v_{j}-E_{l})$ is the passive leak current and  $I_{na}=g_{na} m^3 h (v_{j}-E_{na})$ and  $I_{k}=g_{k} n^4 (v_{j}-E_{k})$ are sodium and potassium currents respectively. $g_{l}=0.3mS/cm^2$ is the conductance for the leak current and $g_{na}=120mS/cm^2$ and $g_{k}=36mS/cm^2$ are the maximum conductance for the sodium and potassium ions, and $E_{l}=10.6mV$, $E_{na}=115mV$ and $E_{k}=-12mV$ are reversal voltages for the leak, sodium and potassium currents respectively. $m_{j}$ $(h_{j})$, activation (inactivation) variable of sodium and $n_{j}$, activation variable of potassium obey the differential equations:
\begin{eqnarray}
\nonumber
\frac{dm_{j}}{dt}=\alpha_{m}(1-m_{j})-\beta_{m}m_{j},\\
\nonumber
\frac{dh_{j}}{dt}=\alpha_{h}(1-h_{j})-\beta_{h}h_{j},\\
\frac{dn_{j}}{dt}=\alpha_{n}(1-n_{j})-\beta_{n}n_{j},
\end{eqnarray}
where $\alpha$ and $\beta$ are functions of membrane voltage as can be found in [16].

With the chemical synapses the synaptic current is described by $I_{ij}=a_{ij} \bar{g}_{ij} s_{ij}(t-\tau) (v_{j}-E_{syn})$ where $\bar{g}_{ij}$ is the synaptic maximum conductivity and $E_{syn}$ is the synaptic reversal potential. $s_{ij}(t)$ is the synaptic activity function defined via:
\begin{equation}
\frac{ds_{ij}}{dt}=\alpha f(v_{i}-v_{th}) (1-s_{ij})-\beta s_{ij},
\end{equation}
with $\alpha$ and $\beta$ defining the activation and deactivation time constants, $v_{th}=20 mV$ is the threshold voltage for the activation of the synapse and $f$ is the threshold function $f(x)=1/2[1+\tanh (5x)]$.

The parameters we have chosen are such that with $I_{ext}=0$, the resting potential of the neuron is zero; so the choice $E_{syn}=80 mV$ is reasonable for excitatory neurons. Inspired by typical time constants of the activation and deactivation of excitatory synapses with AMPA-receptors, we have chosen $\alpha=10$ and $\beta=0.5$ as the activation and deactivation time constants for fast synapses\cite{gerstner}.

\end{document}